%% file: main.tex
\documentclass[aip,graphicx, reprint]{revtex4-1}

\usepackage{graphicx}
\usepackage{amsmath,amssymb}

\newcommand{\angstrom}{\textup{\AA} }

\usepackage{wrapfig}

\draft 

\begin{document}


\title{Structural Dynamics of incommensurate Charge-Density Waves tracked by Ultrafast Low-Energy Electron Diffraction} 



\author{G. Storeck}
\email[]{gero.storeck@uni-goettingen.de}
\affiliation{4th Physical Institute, Solids and Nanostructures, University of Göttingen, Göttingen 37077, Germany}

\author{J. G. Horstmann}
\affiliation{4th Physical Institute, Solids and Nanostructures, University of Göttingen, Göttingen 37077, Germany}

\author{T. Diekmann}
\affiliation{4th Physical Institute, Solids and Nanostructures, University of Göttingen, Göttingen 37077, Germany}

\author{S. Vogelgesang}
\affiliation{4th Physical Institute, Solids and Nanostructures, University of Göttingen, Göttingen 37077, Germany}

\author{G. von Witte}
\affiliation{4th Physical Institute, Solids and Nanostructures, University of Göttingen, Göttingen 37077, Germany}

\author{S. V. Yalunin}
\affiliation{4th Physical Institute, Solids and Nanostructures, University of Göttingen, Göttingen 37077, Germany}

\author{K. Rossnagel}
\affiliation{Institute of Experimental and Applied Physics, Kiel University, 24098 Kiel, Germany}
\affiliation{Ruprecht Haensel Laboratory, Deutsches Elektronen-Synchrotron DESY, 22607 Hamburg, Germany}

\author{C. Ropers}
\email[]{cropers@gwdg.de}
\affiliation{4th Physical Institute, Solids and Nanostructures, University of Göttingen, Göttingen 37077, Germany}
\affiliation{Max Planck Institute for Biophysical Chemistry (MPIBPC), Göttingen, Am Fassberg 11, 37077 Göttingen,
Germany}


\date{\today}

\begin{abstract}

We study the non-equilibrium structural dynamics of the incommensurate and nearly-commensurate charge-density wave phases in 1\textit{T}-TaS\textsubscript{2}. Employing ultrafast low-energy electron diffraction (ULEED) with 1 ps temporal resolution, we investigate the ultrafast quench and recovery of the CDW-coupled periodic lattice distortion. Sequential structural relaxation processes are observed by tracking the intensities of main lattice as well as satellite diffraction peaks as well as the diffuse scattering background. Comparing distinct groups of diffraction peaks, we disentangle the ultrafast quench of the PLD amplitude from phonon-related reductions of the diffraction intensity.
Fluence-dependent relaxation cycles reveal a long-lived partial suppression of the order parameter for up to ~60 picoseconds, far outlasting the initial amplitude recovery and electron-phonon scattering times. This delayed return to a quasi-thermal level is controlled by lattice thermalization and coincides with the population of zone-center acoustic modes, as evidenced by a structured diffuse background. The long-lived non-equilibrium order parameter suppression suggests hot populations of CDW-coupled lattice modes. Finally, a broadening of the superlattice peaks is observed at high fluences, pointing to a nonlinear generation of phase fluctuations.


\end{abstract}


\maketitle 

\section{Introduction}
The spontaneous breaking of a continuous symmetry is a fundamental concept of physics with broad relevance in such diverse areas as particle physics \cite{weinberg_quantum_1995}, cosmology \cite{kibble_topology_1976, zurek_cosmological_1985}, and condensed matter physics \cite{auerbach_interacting_2012, altland_condensed_2010}. An essential consequence of this symmetry breaking is the emergence of new amplitude and phase excitations of the fields considered, exemplified in the Higgs mechanism \cite{higgs_broken_1964} and massless Nambu-Goldstone bosons \cite{nambu_quasi-particles_1960, goldstone_field_1961}, respectively. Moreover, the degenerate ground state of such systems allows for nontrivial topological states, as in the case of magnetic vortices \cite{auerbach_interacting_2012}.

Electron-lattice interaction is an important source of symmetry breaking in solids, most prominently in superconductivity and the formation of charge-density wave (CDW) phases \cite{frohlich_theory_1954, peierls_quantum_1955, bardeen_theory_1957, gruner_density_1994}. Specifically, CDWs constitute a periodic modulation of the charge density by electron-hole pairing \cite{gruner_density_1994}, coupled to a periodic lattice distortion (PLD) \cite{chapman_experimental_1984, minor_phason_1989, pouget_neutron-scattering_1991} and an electronic gap\cite{gweon_direct_1998, uchida_infrared_1981, hirata_temperature_2001, duffey_raman_1976}. The emergence, correlations and fluctuations of symmetry-broken CDW states can be revealed in the time domain by ultrafast measurement techniques. In this way, quenches of the electronic gap coupled to coherent amplitude oscillations\cite{hellmann_ultrafast_2010, perfetti_femtosecond_2008, liu_possible_2013, petersen_clocking_2011, perfetti_time_2006, sohrt_how_2014, demsar_femtosecond_2002}, light-induced PLD dynamics\cite{eichberger_snapshots_2010, erasmus_ultrafast_2012, laulhe_x-ray_2015, han_structural_2012} and phase transitions have been investigated\cite{hellmann_ultrafast_2010, nicholson_beyond_2018,shi_ultrafast_2019}. In particular, ultrafast structural probes trace changes of structural symmetry\cite{haupt_ultrafast_2016, han_ultrafast_2015} and long-range ordering following a phase transformation\cite{vogelgesang_phase_2017, lantz_domain-size_2017}.

However, while the initial quench and coherent amplitude dynamics of CDW systems following short-pulsed excitation are rather well-characterized\cite{hellmann_ultrafast_2010, perfetti_femtosecond_2008, petersen_clocking_2011, perfetti_time_2006, sohrt_how_2014, demsar_femtosecond_2002}, the subsequent path to thermal equilibrium, including the roles of different collective modes in re-establishing a thermal CDW state, are far less understood. In particular, a sensitive structural probe is required to study the interplay of CDW-coupled excitations and regular phonons. 

Here, we employ ultrafast low-energy electron diffraction, a recently developed surface-sensitive structural probe \cite{vogelgesang_phase_2017, gulde_ultrafast_2014, storeck_nanotip-based_2017, horstmann_coherent_2019}, to give a comprehensive account of the non-equilibrium structural dynamics of the incommensurate charge-density wave phases at the surface of 1\textit{T}-TaS\textsubscript{2}. Harnessing the sensitivity of ULEED to the out-of-plane periodic lattice displacements of the sulfur atoms, we isolate the dynamics of an optically-induced amplitude quench from a multi-stage excitation of phonons. Following a rapid partial recovery, we observe a surprisingly long-lived non-thermal amplitude suppression that equilibrates only after approximately 60 ps. Energy transfer to acoustic phonons is required to re-establish a thermal value of the PLD amplitude, suggesting that transient populations of collective CDW modes have a lasting impact on the structural order parameter.

\section{Materials System and Experimental Approach}

\begin{figure*}[ht!]
    \centering
    \includegraphics[width=16cm]{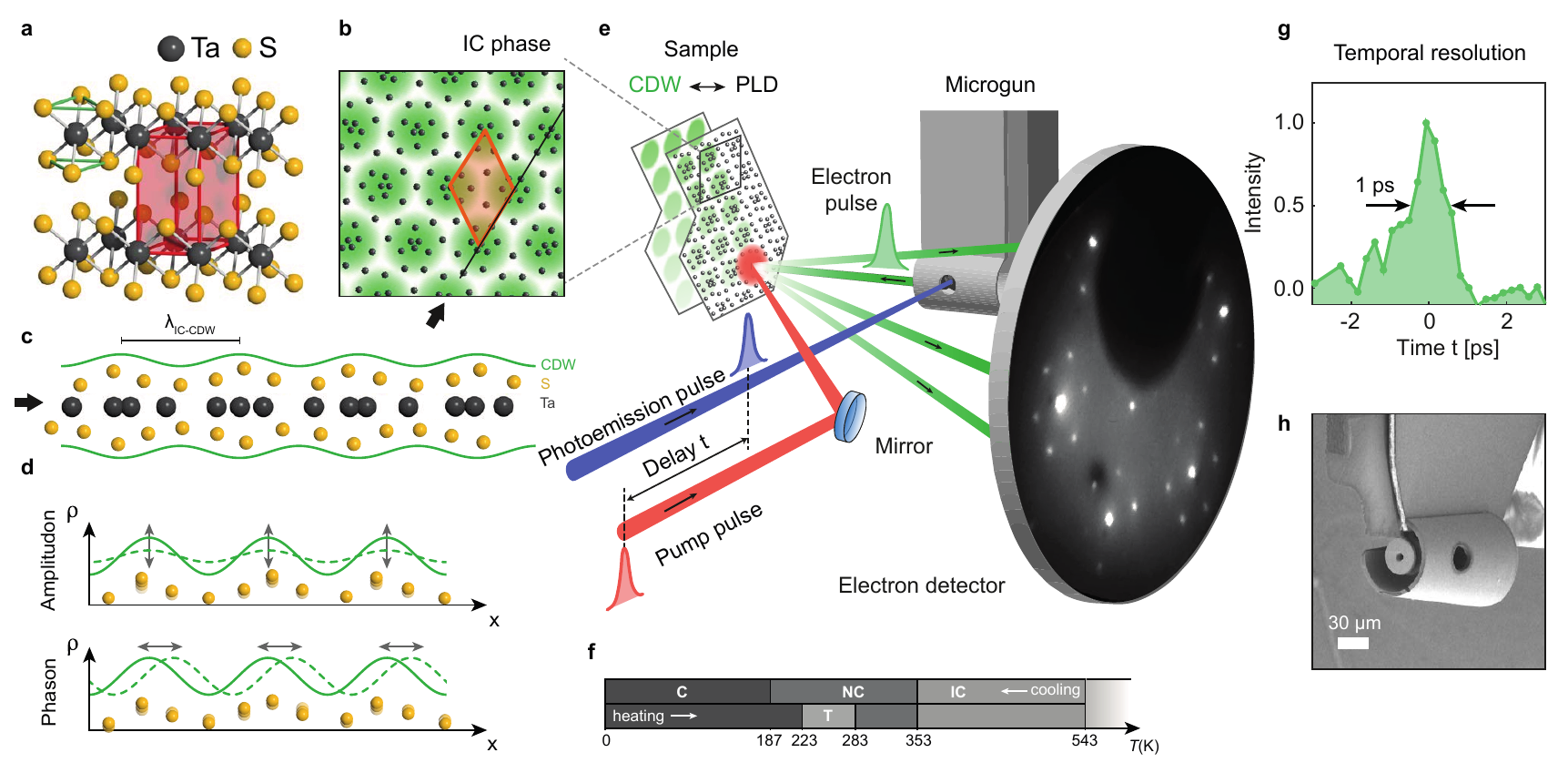}
    \caption{Materials system and experimental setup. (a) Layered transition metal dichalcogenide 1\textit{T}-TaS\textsubscript{2} exhibiting a trigonal crystal structure in the high-temperature phase (green lines: octahedral 1\textit{T}-coordination; red: unit cell). (b) Top view of incommensurate (IC) CDW phase illustrating charge density (green), distorted lattice (black dots: Ta atoms, displacements exaggerated) and superstructure unit cell (orange). (c) Side view of a single S-Ta-S trilayer, illustrating the out-of-plane periodic lattice displacements of the sulfur atoms (exaggerated). (d) 1D sketch of CDW amplitude and phase excitations and corresponding lattice fluctuations. (e) Schematic of the experimental setup, showing ultrafast LEED in a backscattering geometry. Ultrashort electron pulses (green) from a nanofabricated electron gun probe the dynamical evolution of the laser-excited surface structure. (f) Temperature-dependent CDW phases. (g) Achieved electron pulse duration of 1 ps (see Appendix \ref{sec:methods} for details). (h) Scanning electron micrograph of miniaturized electron gun.}
    \label{fig:Figure01_SDY}
\end{figure*}

In this work, we study one of the most prominent CDW systems, 1\textit{T}-TaS\textsubscript{2}, which is part of the class of transition metal dichalcogenides. The atomic structure of this material consists of weakly interacting S-Ta-S trilayers\cite{scruby_role_1975, spijkerman_x-ray_1997}, in which the tantalum atoms are octahedrally coordinated between the sulfur atoms (Fig.~\ref{fig:Figure01_SDY}a). This compound has attracted much attention for its various CDW phases\cite{gruner_density_1994, scruby_role_1975, spijkerman_x-ray_1997, vaskivskyi_controlling_2015}, excitations\cite{perfetti_femtosecond_2008, perfetti_time_2006, sohrt_how_2014, demsar_femtosecond_2002, eichberger_snapshots_2010, hellmann_time-domain_2012} (Fig.~\ref{fig:Figure01_SDY}c), and correlation effects\cite{fazekas_electrical_1979, darancet_three-dimensional_2014, law_1t-tas2_2017}, serving as a model system to study, for example, Peierls- versus Mott-type metal-insulator transitions\cite{rossnagel_origin_2011, petersen_clocking_2011}, pressure-induced superconductivity in coexistence with CDWs\cite{sipos_mott_2008}, transitions to metastable 'hidden' CDW states\cite{stojchevska_ultrafast_2014, shi_ultrafast_2019}, the emergence of complex orbital textures \cite{ritschel_orbital_2015}, or quantum spin liquid behavior\cite{klanjsek_high-temperature_2017}. 

The material exhibits multiple temperature-dependent phases (Fig.~\ref{fig:Figure01_SDY}f) with characteristic lattice deformations coupled to electronic structure changes\cite{scruby_role_1975, rossnagel_origin_2011, dardel_spectroscopic_1992}. Starting from a metallic phase with an undistorted trigonal structure (Fig.~\ref{fig:Figure01_SDY}a) above 543~K, the system undergoes a sequence of CDW transitions, forming a commensurate (C) (Mott-insulating) state below 187~K. At intermediate temperatures, two incommensurate phases are found, namely the so-called 'nearly-commensurate' (NC) phase (187-353~K), exhibiting commensurate patches separated by discommensurations \cite{spijkerman_x-ray_1997, nakanishi_nearly_1977, nakanishi_domain-like_1977, ishiguro_high-resolution_1995}, and a homogeneous, fully incommensurate (IC) structure (Fig.~\ref{fig:Figure01_SDY}b) between 353~K and 543~K. The periodic lattice distortions in these phases are characterized by primarily in-plane and out-of-plane displacements of the tantalum and sulfur atoms, respectively (Figs.1b, c). Ultrafast transitions between and manipulation of these phases, as well as their collective modes (Fig.~\ref{fig:Figure01_SDY}d) have been observed in various diffraction and spectroscopy studies \cite{hellmann_ultrafast_2010, perfetti_femtosecond_2008, petersen_clocking_2011, perfetti_time_2006, sohrt_how_2014, demsar_femtosecond_2002, eichberger_snapshots_2010, haupt_ultrafast_2016, vogelgesang_phase_2017, hellmann_time-domain_2012, laulhe_ultrafast_2017, le_guyader_stacking_2017, avigo_excitation_2019, zong_evidence_2018, kusar_anharmonic_2011, mann_probing_2016}.


In our experiments, we employ pulses of electrons at low energies, typically in the range of 40-150~eV, to probe the structural evolution of the NC and IC states in backscattering diffraction. Ulrafast low-energy electron diffraction \cite{vogelgesang_phase_2017, gulde_ultrafast_2014, storeck_nanotip-based_2017, horstmann_coherent_2019} allows us to trace the changes of the diffraction pattern in the time domain, following intense fs-laser illumination (red pulse in Fig.~\ref{fig:Figure01_SDY}e). In this optical-pump/electron-probe scheme, excitation and relaxation processes are sampled by varying the time delay $t$ between the optical pump pulse (red) and the photoemission pulse (blue) generating the electron probe (green). Reducing electron pulse broadening by short propagation lengths, a miniaturized electron gun (Fig.~\ref{fig:Figure01_SDY}h) \cite{storeck_nanotip-based_2017} allows for a temporal resolution of 1 ps (Fig.~\ref{fig:Figure01_SDY}g). Further experimental details are provided in Appendix \ref{sec:methods} (Fig.~\ref{fig:Figure07_SDY}).

\begin{figure*}[ht!]
    \centering
    \includegraphics[width=16cm]{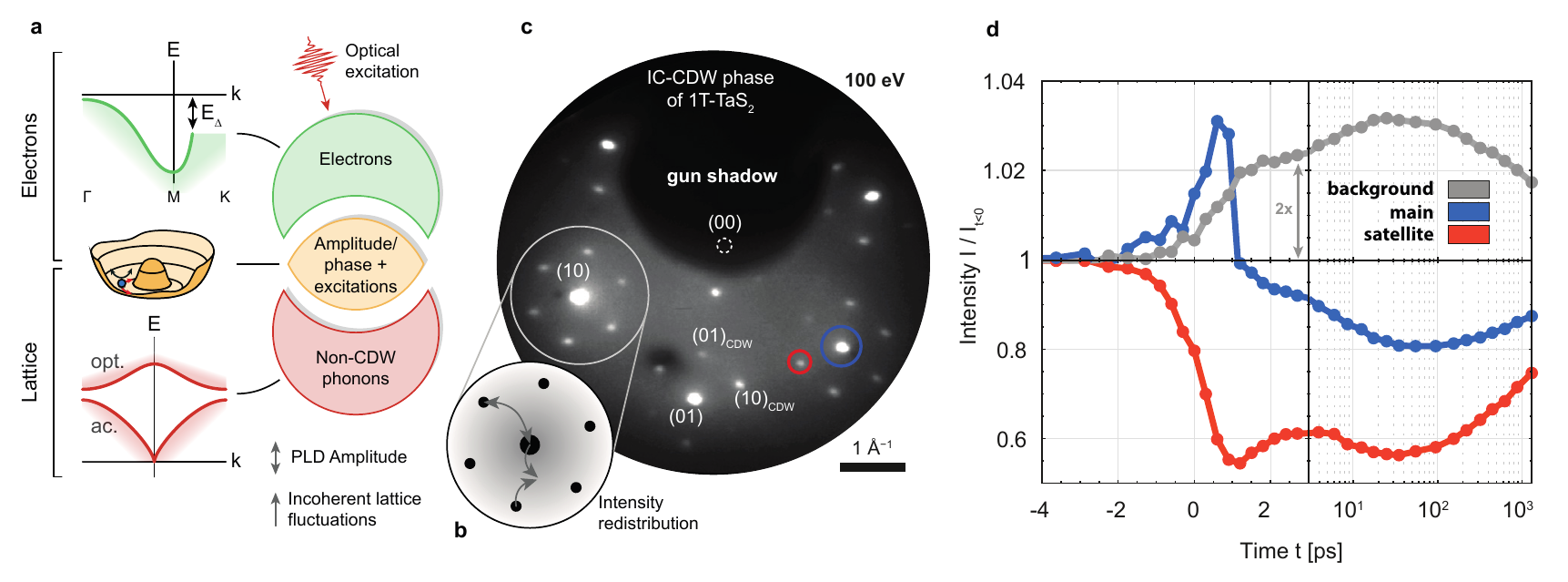}
    \caption{Dynamics and excitations in CDW systems influencing diffraction. (a) Electron and lattice subsystems (right) governing CDW dynamics. Gapped band structure (top, left), symmetry broken CDW state with phase and amplitude excitations (middle), and non-CDW phonons (bottom). (b) Changes in average amplitude and all lattice excitations (CDW and non-CDW) lead to a redistribution of intensity in the electron diffraction pattern. (c) Diffraction pattern of the IC phase of 1\textit{T}-TaS\textsubscript{2} showing main lattice reflexes and first-order PLD-induced satellites (integration time: 90 s, electron energy: 100~eV).  (d) Time-dependent measurement of reflexes (blue and red circles in (c)) and diffuse background (fluence $F=2.5$ mJ/cm$^2$). The main lattice signal is averaged over (10) and (-1 1) spots (blue), the satellite signal over several reflexes. Curves are normalized to the signal at negative times.}
    \label{fig:Figure02_SDY}
\end{figure*}

To facilitate the discussion, we focus the presentation on the response of the IC phase, which has not been studied by ultrafast diffraction, and provide a comprehensive data set of related observations for the NC phase in Appendix \ref{sec:Data for nearly-commensurate (NC) phase}.

The IC phase exhibits a triple-Q CDW/PLD, with lattice displacements for each unit-cell atom of the form \cite{mcmillan_landau_1975, mcmillan_theory_1976}

\begin{equation}
    \textbf{u}(\textbf{L})=\sum_{i=1,2,3} \textbf{A}_i \sin(\textbf{Q}_i\cdot\textbf{L}+\varphi_i)
\end{equation}

for lattice sites $\textbf{L}$, CDW wavevectors $\textbf{Q}_i$ and phases $\varphi_i$. The CDW/PLD texture of a 'dot-lattice' arises for the phasing condition $\sum_i \varphi_i=0$, and for symmetry reasons, the individual plane wave components share a common amplitude $A=|\textbf{A}_i|$. The PLD at a wavelength $\lambda_{IC}=3.53 a$ ($a$: lattice constant) leads to characteristic arrangements of satellite peaks\cite{overhauser_observability_1971, giuliani_structure_1981} around the main lattice diffraction spots, seen in the ULEED pattern displayed in Fig.~\ref{fig:Figure02_SDY}c. As the IC state wave vectors are collinear to the lattice vectors, the satellites are located on the lines connecting the main reflexes. Due to the harmonic (and weak) structural modulation\cite{nakanishi_nearly_1977, nakanishi_domain-like_1977}, only first-order satellites are observed, with an intensity\cite{overhauser_observability_1971} $I_{\text{sat}}\sim |J_1({\textbf{s}}\cdot{\textbf{A}_i})|^2\sim A^2$ ($\textbf{s}$: scattering vector). We note that in this energy range, LEED is a very efficient structural probe of the PLD, because (i) backscattering diffraction is dominated by the sulfur sublattice, and (ii) the large out-of-plane momentum transfer enhances the sensitivity to out-of-plane displacements.

We study the excitation and relaxation of the IC and NC phases, without driving the system across a phase transition\cite{laulhe_x-ray_2015, haupt_ultrafast_2016, vogelgesang_phase_2017, lantz_domain-size_2017, nakanishi_nearly_1977, mcmillan_landau_1975}. The dynamics of this incommensurate Peierls system can be discussed based on a simplified picture of three coupled subsystems, namely, the electronic system exhibiting a gapped band structure (Fig.~\ref{fig:Figure02_SDY}a, top), the collective amplitude and phase excitations around the symmetry-broken CDW ground state (center)\cite{gruner_density_1994}, and the ordinary lattice modes far from the CDW wavevector in reciprocal space, i.e., regular phonons (bottom). 

It is widely established that electronic excitation by an ultrashort laser pulse induces a carrier population above the band gap, which results in a quench of the CDW/PLD amplitude that recovers upon carrier cooling by electron-phonon scattering \cite{perfetti_time_2006, eichberger_snapshots_2010, hellmann_time-domain_2012}. The corresponding sequence of relaxation processes involving the three subsystems causes characteristic changes to the diffraction intensities of the satellite peaks and the main peaks (intensity $I_{\text{main}}$). Specifically, for small PLD amplitudes, the peak intensities are expected to scale as \cite{overhauser_observability_1971, giuliani_structure_1981, wang_thermal-diffuse_1989, ichimiya_reflection_2004}:
\begin{align}
&I_{\text{sat}}\sim A^2\,e^{-2W_\varphi}\,e^{-2W_\textbf{s}}, \label{appa}
\\
&I_{\text{main}}\sim (1-c_\textbf{s}A^2)\ e^{-2W_\textbf{s}}, \label{appb}
\end{align}

These expressions reflect that a light-induced quench of the mean PLD amplitude $A$  will lead to a redistribution of intensity from the satellites to the main peaks \cite{eichberger_snapshots_2010, erasmus_ultrafast_2012, li_ultrafast_2019}. Different main reflexes are sensitive to the PLD to a varying degree, which requires the introduction of the factor $c_\textbf{s}$ that depends on the momentum transfer \textbf{s}. Inelastic scattering by generated phonons transfers intensity from the reflexes to a diffuse background (Fig.~\ref{fig:Figure02_SDY}b) \cite{van_hove_low-energy_1986, stern_mapping_2018, waldecker_momentum-resolved_2017, otto_ultrafast_2019}, leading to a peak suppression by a Debye-Waller factor $\exp(-2W_\textbf{s})$ \cite{ichimiya_reflection_2004,van_hove_low-energy_1986}. The general form of the exponent\cite{ichimiya_reflection_2004, van_hove_low-energy_1986} $W_\textbf{s}\sim \sum_{ph}(\textbf{s}\cdot \textbf{u}_{ph})^2$ sums over the momentum transfer projected onto phonon displacements $\textbf{u}_{ph}$ in various branches. According to Overhauser\cite{overhauser_observability_1971}, phase fluctuations result in the additional 'phason Debye-Waller factor' $e^{-2W_\varphi}=e^{-{\langle\varphi^2\rangle}}$, which only affects the satellite spots and also causes diffuse scattering in the vicinity of the satellite peaks \cite{wang_thermal-diffuse_1989, lee_phase_2012}.
Finally, dislocation-type topological defects in the CDW may broaden the superlattice peaks and also reduce the PLD in the dislocation core \cite{vogelgesang_phase_2017, zong_evidence_2018}.

\section{Results and Analysis}

Our ULEED experiments directly show the characteristic diffraction changes mentioned above: In the exemplary data displayed in Fig.~\ref{fig:Figure02_SDY}d, a main lattice peak (blue) exhibits a transient intensity increase after the pump pulse, before experiencing an initially rapid and then slowed suppression to a minimum at $t=60$ ps. The satellite peaks (red), on the other hand, are first suppressed, before approaching a similar trend as the main peak beyond approximately 10 ps. Both the satellite and main peak intensities are significantly reduced by phonon populations\cite{van_hove_low-energy_1986}. These are evident in the diffuse background (gray), which mirrors the suppression of the reflexes, with a step-like increase in the first ps and a slower rise to a maximum at the delay of 60 ps. The initial step can be interpreted as the excitation of a broad population of optical and acoustic phonons on the timescale of electron-phonon energy relaxation ($<1$ ps)\cite{demsar_femtosecond_2002}, while the slower timescale corresponds to phonon-phonon equilibration\cite{gu_phonon_2014} and the population of low-energy acoustic modes. LEED intensities are rather sensitive to the large amplitudes of low-frequency modes, particularly those with out-of-plane polarization. Specifically, phonon modes with out-of-plane displacements $\textbf{u}_{ph}$ have a more pronounced Debye-Waller factor due to the backscattering geometry with a primarily out-of-plane scattering vector of the electron. In addition, these modes exhibit comparatively slow phase velocities, as is typical for layered van-der-Waals materials\cite{mohr_phonon_2007}. Thus, the prominent main lattice suppression evolving over tens of picoseconds primarily stems from the increasing population of low frequency acoustic modes modulating the layer distance.

\begin{figure}
    \centering
    \includegraphics{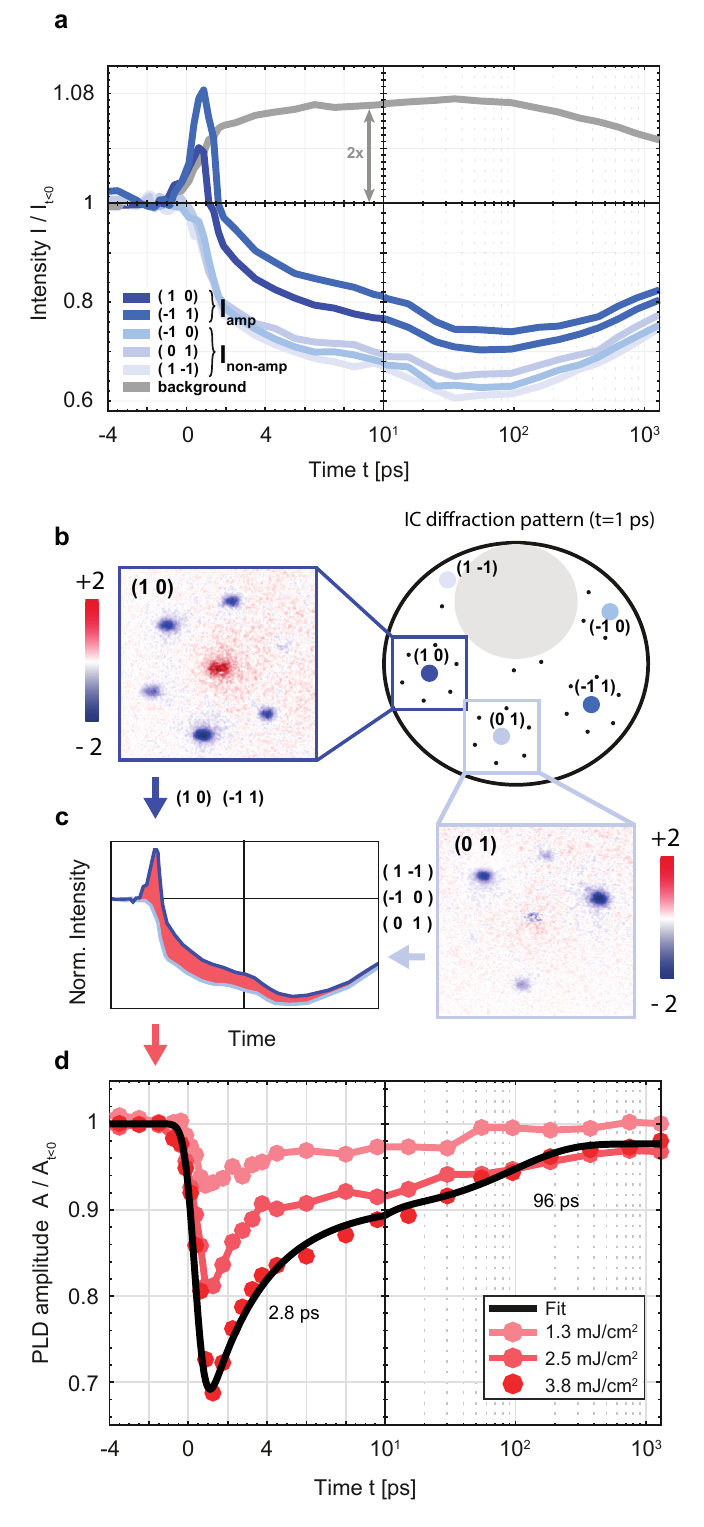}
    \caption{Amplitude dynamics of the PLD obtained from main lattice reflexes. (a) Time-dependent intensity of visible main lattice reflexes and integrated background intensity, for a fluence of ${F\,=\,3.8}$\,mJ/cm$^2$. Two inequivalent classes of spot groups are found, featuring a strong (dark blue) and weak (light blue) sensitivity to the amplitude quench. (b) Sketch of IC diffraction pattern, and parts of the difference diffraction image ($I_{t=1\text{ps}} - I_{t<0}$) around the (10) and (01) main reflexes (insets). (c) Schematic comparison of peak intensities in the spot groups. Red area highlights different sensitivities to the PLD. (d) Extracted PLD amplitude quench and relaxation (see also Appendix \ref{sec:methods}) for three fluences, showing a rapid and a slower relaxation component. (Time constants from a biexponential fit (black line) to the highest fluence data: (2.8 $\pm$ 0.3) ps and (96 $\pm$ 3) ps).}
    \label{fig:Figure03_SDY}
\end{figure}


These strong Debye-Waller factors complicate an analysis of the temporal evolution of the amplitude quench. On the other hand, our experimental data shows that the different reflexes share a common phonon-induced peak suppression. In the following sections \ref{subsectionA} and \ref{subsectionB}, we pursue two approaches of disentangling the dynamics of the structural order parameter from the phonon population, exploiting (Section~\ref{subsectionA}) the different sensitivities of two inequivalent classes of main lattice reflections to the PLD, and (Section~\ref{subsectionB}) the direct sensitivity of the satellite reflexes to the PLD.

\subsection{Amplitude Analysis based on Main Lattice Reflexes}
\label{subsectionA}

Concerning the time-dependent peak intensity, the main reflexes fall into two different groups. Whereas all five visible main peaks show a suppression opposite to the increase in diffuse background (Fig.~\ref{fig:Figure03_SDY}a), we find that the transient amplitude signal is prominent only in the (1 0) and (-1 1) peaks, while it is largely absent in the (0 1), (-1 0), and (1 -1) peaks (see also difference maps in Fig.~\ref{fig:Figure03_SDY}b)\cite{noauthor_note_nodate}. These two groups of peaks are crystallographically distinct, and the peaks within each group are equivalent in the effective threefold symmetry of the 1T structure \cite{von_witte_surface_2019}. The different sensitivity of the peak intensities to the PLD is a particular feature of LEED, as described in the following.

In the electron energy range of 70-110~eV, diffraction intensities are mainly governed by scattering from sulfur atoms, due to large atomic scattering factors\cite{spijkerman_x-ray_1997, von_witte_surface_2019}. As a result of the CDW-induced contraction of the tantalum sub-lattice, the sulfur atoms predominantly exhibit out-of-plane displacements. In backscattering, the opposing directions for the displacements in the upper and lower sulfur layers within each S-Ta-S trilayer\cite{spijkerman_x-ray_1997, von_witte_surface_2019} (Fig.~\ref{fig:Figure01_SDY}c) lead to an interference with enhanced or suppressed sensitivity of the two groups of main lattice peaks to the lattice distortion. This feature is expected in all CDW phases of 1\textit{T}-TaS\textsubscript{2}, which share the phasing condition mentioned above (compare Fig.~\ref{fig:Figure01_SDY}b). Experimentally, we found the same trend in experiments on the NC phase (see Appendix \ref{sec:Data for nearly-commensurate (NC) phase}), which exhibits different wavevectors but the same phasing between the three CDWs. In order to further corroborate these findings and considering the importance of multiple scattering in LEED, we conducted dynamical LEED simulations for a PLD of a varying amplitude and as a function of the electron beam energy (see Appendix \ref{sec:Dynamical LEED computation}). In these simulations, for computational reasons, the commensurate modulation was employed, taking quantitative displacements from a recent LEED reconstruction\cite{von_witte_surface_2019}. Importantly, the dynamical LEED simulations qualitatively reproduce our experimental findings of different sensitivities to the PLD by the two groups of main lattice peaks. Moreover, the simulations predict an energy-dependent and strongly reduced PLD-sensitivity at an electron energy of 80~eV. Indeed, experiments at this lower energy show that the transient increase of the main peak is generally much weaker (see additional data in Appendix \ref{sec:Data at 80eV electron energy}).

We employ these different sensitivities to the PLD to derive a phonon-corrected amplitude signal. Specifically, we remove the phonon-induced Debye-Waller suppression by normalizing the intensity of the PLD-sensitive peaks to that of the  weakly sensitive peaks (Fig.~\ref{fig:Figure03_SDY}c; see Appendix \ref{sec:methods} for details). The resulting phonon-corrected amplitude suppression is displayed in Fig.~\ref{fig:Figure03_SDY}d for three pump fluences. In each case, the amplitude exhibits a rapid initial quench (within our temporal resolution), and a recovery with an exponential time constant of about 3 ps. The re-establishment of the amplitude is, however, incomplete, slowing down considerably beyond 4 ps, and lasting well into the range of tens to one-hundred picoseconds.

\subsection{Amplitude Analysis based on Satellite Reflexes}
\label{subsectionB}

\begin{figure}[ht!]
    \centering
    \includegraphics[width=7cm]{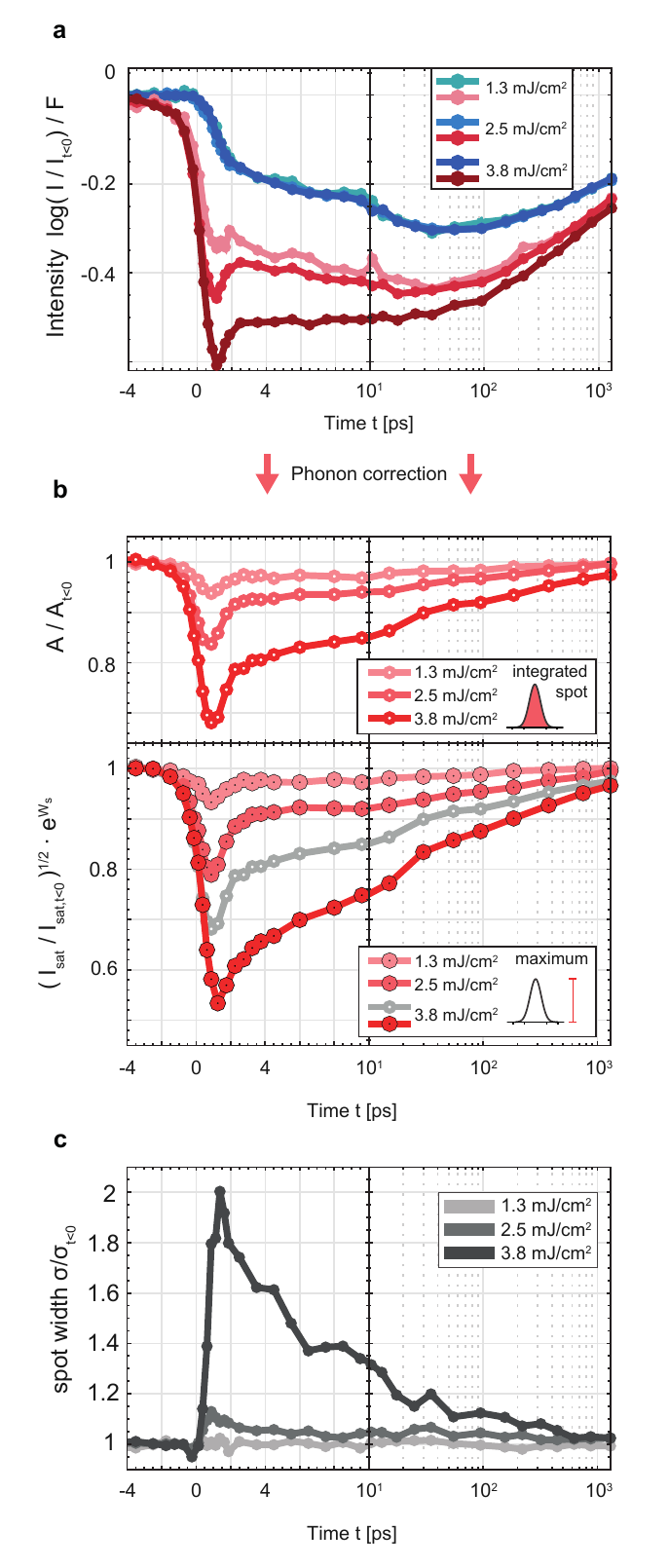}
    \caption{Amplitude  dynamics  of the PLD  obtained  from  satellite reflexes. (a) Logarithm of normalized main lattice and satellite peak intensities (mean value), divided by fluence, versus time delay. While the main peak intensities (blue) collapse to a single curve due to the exponential (in fluence) Debye-Waller-type suppression, the satellite intensities (red) show a strong fluence-dependent behavior for early times, before converging for long time delays. (b) Phonon-corrected PLD amplitude obtained from integrated (top) and maximum (bottom) satellite intensities. (c) Fluence-dependent
    azimuthal spot width $\sigma$ versus time.}
    \label{fig:Figure04_SDY}
\end{figure}

We now aim at characterizing the evolution of the mean amplitude based on the satellite peak intensities, again removing a time-dependent phonon Debye-Waller factor. To this end, we compare the intensities of the main peaks with weak PLD-sensitivity to the satellite peaks. In Fig.~\ref{fig:Figure04_SDY}a, we plot the logarithm of the these intensities (normalized to the signal at $t<0$), divided by the fluence. For all three fluences, the traces of the main lattice peaks collapse to a single universal curve (blue), illustrating the phonon-induced Debye-Waller suppression $W_{\textbf{s}}$ and its proportionality to fluence. The satellite peaks show a non-exponential fluence dependency in their suppression and recovery. At low fluences, however, where only a minor amplitude quench is induced, the satellite peak suppression closely follows that of the main peaks. We use this information to derive a phonon-corrected amplitude signal from the satellite peaks (see Appendix \ref{sec:methods}). Figure \ref{fig:Figure04_SDY}b shows the resulting amplitude evolution. For this graph, the satellite intensities were integrated over circular masks in the diffraction pattern (width of $\Delta k_{\text{sat}}=0.36 \ \angstrom^{-1}$), therefore including also electrons scattered by a small angle from the reflex. We find a very similar behavior as from the main peak analysis (see Section~\ref{subsectionA}), namely a rapid and fluence-dependent quench, a fast initial recovery and a rather persistent partial suppression, and we therefore consider this quantity as representative for the evolution of the amplitude~$A$.

A somewhat different curve is obtained by utilizing not the area-integrated intensity, but the maximum intensity on top of the diffraction spot (bottom graph in Fig.~\ref{fig:Figure04_SDY}b). Whereas the maximum and integrated intensities behave similarly at low fluence, at the highest fluence, the suppression of the maximum intensity exceeds that of the integrated intensity (grey curve from integrated intensity shown again for comparison). Moreover, the recovery of the maximum proceeds more gradually than the integrated intensity.

The difference between the evolution of the integrated and maximum intensities implies a change in diffraction peak shape, which is analyzed in Fig.~\ref{fig:Figure04_SDY}c. Plotting the azimuthal width of the diffraction peak, we find a significant time-dependent broadening for the highest fluence.

This effective broadening may be a result of several phenomena: (i) Diffuse scattering to the wings of the peak by low-energy phase excitations \cite{minor_phason_1989} will suppress the reflex maximum via the phason Debye-Waller factor $\exp(-2W_{\varphi})$ while largely maintaining the integrated intensity. (ii) An overall peak broadening from reduced correlation lengths will arise from the generation of CDW dislocation-type topological defects \cite{vogelgesang_phase_2017, zong_evidence_2018}.
Except for the amplitude suppression in the dislocation core, this broadening also preserves the integrated intensity. At this point, we cannot rule out either scenario, and a more detailed spot profile analysis or higher momentum resolution may be required to further elucidate the different contributions.

\subsection{Non-equilibrium dynamics of the amplitude}
\label{subsectionC}

\begin{figure}[!]
    \centering
    \includegraphics{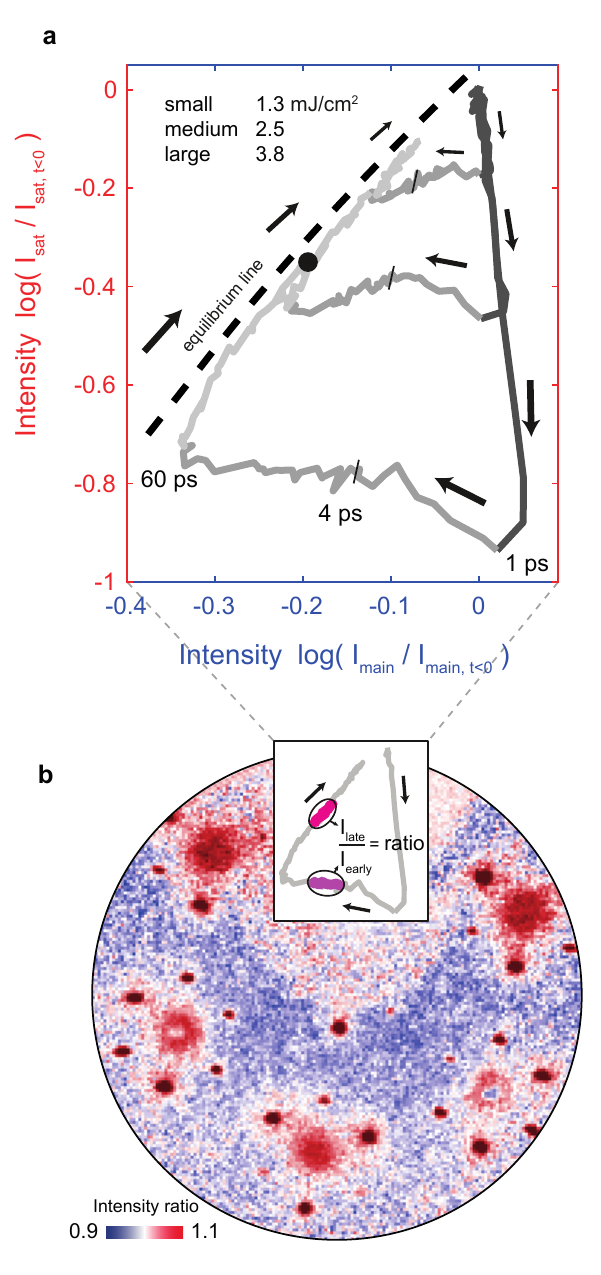}
    \caption{Path to equilibrium. (a) Intensities of satellite and main peaks (with PLD sensitivity) plotted against each other, leading to cyclic trajectories in a 2D plane with varying size. Note that all curves reach a common equilibrium line after approximately 60 ps. The gray color scale highlights certain time intervals (dark gray: 0-1 ps, intermed. gray: 1-60 ps, light gray: 60-1500 ps). The same combination of intensity suppressions is found for different fluences at different times (black circle corresponds to 1500 ps/290 ps at high/intermediate fluence). (b) Ratio of time-integrated frames exhibits prominent pedestals around diffraction peaks, pointing to an enhanced acoustic phonon population on the equilibrium line. Late frames (dark magenta in inset, $t=790 \ldots 1500$ ps) divided by early frames (light magenta in inset, $t=4.5 \ldots 10$ ps).}
    \label{fig:Figure05_SDY}
\end{figure}

The incomplete recovery and persistent suppression of the PLD amplitude, independently obtained from the main (Fig.~\ref{fig:Figure03_SDY}d) and satellite (Fig.~\ref{fig:Figure04_SDY}b) reflexes, warrants further investigation. It implies that the system is either thermalized at a higher temperature with reduced equilibrium amplitude\cite{erasmus_ultrafast_2012}, or, alternatively, that non-equilibrium dynamics inhibit the recovery of the order parameter. It was previously suggested for the NC phase that the rapid recovery results in a thermalized system at elevated temperature\cite{eichberger_snapshots_2010}. Specifically, this would entail equilibrium between the electronic and different structural degrees of freedom after approximately 4 ps. 

\begin{figure*}[ht!]
    \centering 
    \includegraphics{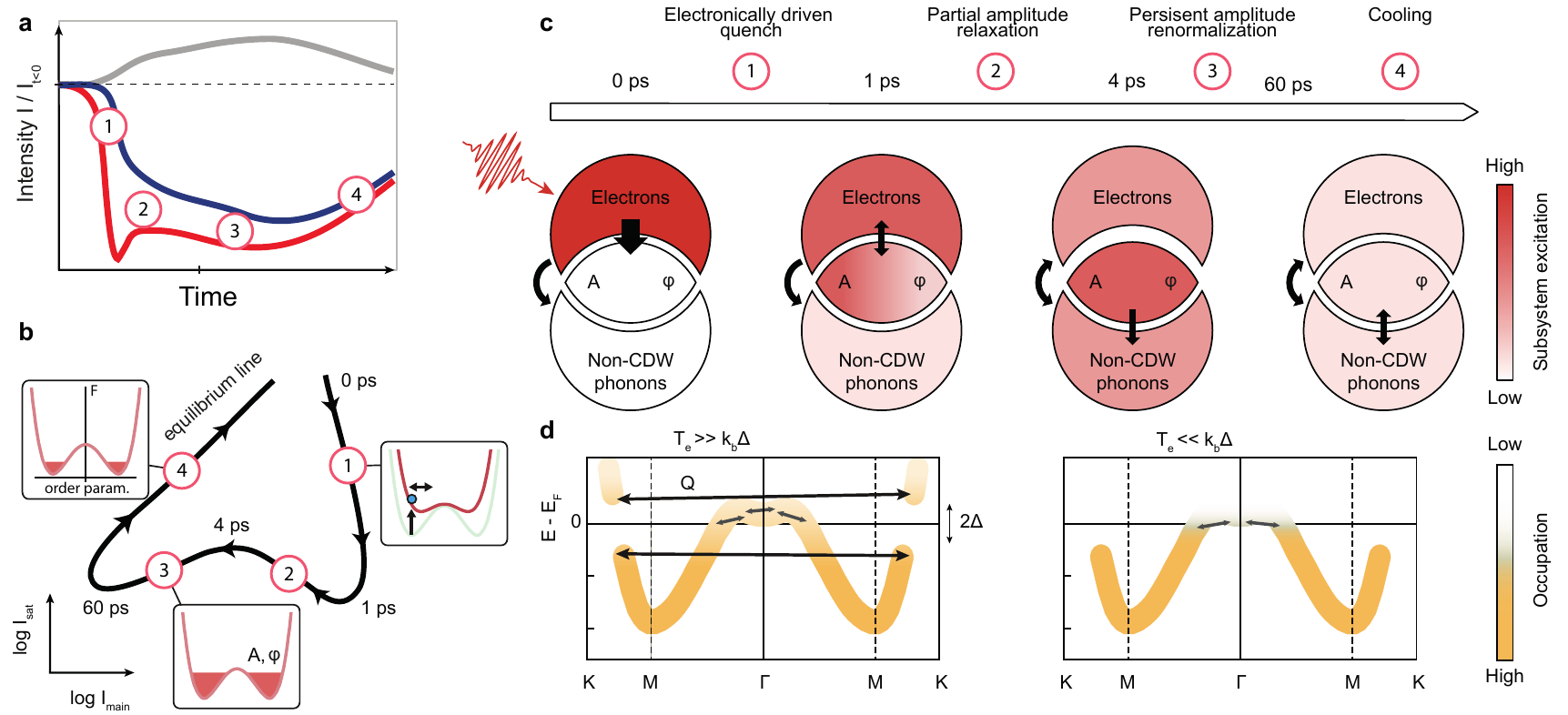}
    \caption{Linking relaxation pathways to CDW/PLD dynamics. (a) and (b) are simplified sketches of Figs. 2d and 4a, respectively, highlighting four phases of the relaxation process observed in the data. (c) Illustration of sequential relaxation process and the excitation levels of the three subsystems. The color shade represents energy content/temperature, black arrows indicate energy flow. (d) Simplified electronic band structure and populations (saturation of orange line) for high (left) and low (right) electronic temperature. Arrows indicate electron-lattice scattering processes. Scattering between gap regions (momentum transfer Q) is suppressed for reduced electronic temperatures.}
    \label{fig:Figure06_SDY}
\end{figure*}

As shown in the following, we have evidence for a sustained non-thermal suppression of the order parameter. In Fig.~\ref{fig:Figure05_SDY}, we consider in more detail the path to thermal equilibrium. An instructive depiction is obtained by plotting the main and satellite intensities against each other, resulting in cyclic trajectories in a two-dimensional plane (Fig.~\ref{fig:Figure05_SDY}a), traced out over time in a clockwise fashion. At long delays (beyond 100 ps), the curves for all fluences follow a universal path (dashed line) representing a thermalized system at elevated temperatures, cooling down. Different trajectories reach the same combination of intensities at different times. For instance, the high-fluence trajectory exhibits the same combination of intensity suppressions at 1500 ps as the intermediate fluence at a somewhat earlier time of 290 ps (black circle in Fig.~\ref{fig:Figure05_SDY}a). Once the trajectory reaches this line, the surface is in local thermal equilibrium, characterized by a single temperature, and the satellite peak suppression is composed of a Debye-Waller factor as well as a thermal reduction of the amplitude. The further progression of the system, i.e. its cooling, is governed by thermal diffusion to the bulk.

All points \textit{displaced} from the dashed line represent deviations from a thermal state, with the distance being a very sensitive measure of the structural non-equilibrium. For example, within the first picosecond after the excitation (dark segments of the curves), the rapid quench of the order parameter causes a reduction of satellite intensity and a moderate enhancement of the main lattice signal, with a fluence-dependent maximum displacement from thermal equilibrium (corresponding curves for the main peaks insensitive to the amplitude are found in Appendix \ref{sec:Relaxation cycles for main peaks}). The recovery to the thermal state now proceeds through various stages and in a fluence-dependent manner. After about 4 ps (see marks), the fast component of the amplitude recovery is completed (cf. Fig.~\ref{fig:Figure03_SDY}d, compare also Ref. 74)\cite{eichberger_probing_2014}. However, the system remains far from the equilibrium state, i.e., exhibits a lower-than-thermal satellite intensity. Interestingly, for all curves, a surprisingly long time of approximately 60 ps is required to reach the thermal state. This depiction directly shows that the persistent amplitude suppression discussed in Fig.~\ref{fig:Figure03_SDY}d and \ref{fig:Figure04_SDY}b is in fact not of a thermal nature, and that we have a pronounced deviation from equilibrium between the degrees of freedom affecting the diffraction intensities.

To identify the origin of this long-lived amplitude suppression, we first note that the time at which the system reaches a thermal amplitude nearly coincides with the strongest suppression of the main lattice peaks. As this time also corresponds to the maximum intensity of the diffuse background (cf. Figs. \ref{fig:Figure02_SDY}d, \ref{fig:Figure03_SDY}a), the full equilibration of lattice fluctuations appears to be critical in controlling the structural order parameter. In particular, this lattice equilibration induces a significant increase of diffuse background intensity around main lattice peaks (difference image in Fig.~\ref{fig:Figure05_SDY}b), directly pointing to the excitation of low-energy acoustic modes near the center of the Brillouin zone.

\section{Discussion}

Taken together, these observations indicate the sequence of relaxation processes illustrated in Fig.~\ref{fig:Figure06_SDY}b, which can be related to the intensity curves (Fig.~\ref{fig:Figure04_SDY}a) and the cyclic trajectories introduced above (simplified sketch in Fig.~\ref{fig:Figure06_SDY}a). Within the first picosecond, optical excitation of the electronic system leads to a CDW amplitude quench and a strong deformation of the potential energy landscape (see insets in Fig.~\ref{fig:Figure06_SDY}b), which triggers cooperative motion of the lattice towards its unmodulated state, including the excitation of coherent amplitude modes (stage 1) \cite{demsar_femtosecond_2002, hellmann_time-domain_2012}. Facilitated by the generation of high-energy lattice modes, the electron system cools down within few ps (stage 2), and as a result, the electronic potential and amplitude partially recover. The remaining PLD suppression in the following stage 3 strongly indicates a substantial population of CDW-coupled lattice excitations (Fig.~\ref{fig:Figure06_SDY}b, red filling in bottom inset), such as amplitudons, phasons, and possible dislocation-type topological defects. Remaining non-thermal electronic excitations, on the other hand, can be largely ruled out at these late times, based on results from time- and angle-resolved photoemission spectroscopy \cite{perfetti_time_2006, petersen_clocking_2011, hellmann_time-domain_2012, ligges_ultrafast_2018, shi_ultrafast_2019}. While it is known that phonon equilibration may take tens of picoseconds\cite{trigo_fourier-transform_2013, harb_phonon-phonon_2016}, the present observations are significant in the sense that the persistent structural non-equilibrium is found to directly lead to an amplitude suppression via long-lived CDW-coupled excitations.

Both amplitude and phase modes are expected to be rather efficiently excited by the optical pump, either directly by the deformation of the electronic potential (amplitude modes) \cite{sohrt_how_2014} or by electron-lattice scattering between gap regions (Fig.~\ref{fig:Figure06_SDY}d). Specifically, Fermi surface nesting is expected to result in a high probability of scattering events with a momentum transfer around the CDW wavevector $Q$ (Fig.~\ref{fig:Figure06_SDY}d, left). Subsequent cooling of the carrier temperature below the energy scale of the electronic gap will effectively suppress these inelastic scattering pathways (Fig.~\ref{fig:Figure06_SDY}d right) and decouple the subsystems (Fig.~\ref{fig:Figure06_SDY}c), contributing to the persistent amplitude suppression in stage (3). Full lattice thermalization and the excitation of zone-center acoustic modes is then only achieved after 60 ps, from which point on the equilibrated system cools down (stage 4).


Let us consider the possible roles of different CDW excitations in the long-lived amplitude suppression, namely amplitudons, phasons, and CDW dislocation defects. Spatiotemporal variations of the amplitude and phase affect the observable value of $A$. Specifically, amplitudons represent amplitude oscillations $\Delta A$ around an equilibrium amplitude $A_0$, leading to an observed average value of $\langle A_0+\Delta A \rangle$. By an anharmonicity of the electronic potential, these oscillations become asymmetric, and a high population of amplitudons can reduce the value of $A$. In the case of phasons, despite early theoretical and experimental work\cite{chapman_experimental_1984, minor_phason_1989, overhauser_observability_1971, krivoglaz_diffuse_1996, giuliani_structure_1981, axe_debye-waller_1980, adlhart_dynamic_1982,aslanyan_debye-waller_1998, aslanyan_comment_2005}, a unifying picture has not been reached, and recent assignments of their contribution in diffraction studies range from largely negligible\cite{li_ultrafast_2019} to dominant\cite{lee_phase_2012}. While our results do not definitely resolve this issue, the redistribution of scattering intensity near the satellite peaks suggests significant spatial or spatiotemporal phase distortions.

CDW dislocation defects should also be considered as a possible cause for the long-lived order parameter suppression, as they have been observed as a consequence of phase transitions, e.g. in 1\textit{T}-TaS\textsubscript{2} \cite{vogelgesang_phase_2017} or LaTe$_3$ \cite{zong_evidence_2018, Kogar2020}. The fact that we find a significant spot broadening of the satellites (Fig.~\ref{fig:Figure04_SDY}c) most strongly at high fluences suggests a nonlinear dependence of phase fluctuations. This would be consistent with either CDW dislocations generated by critical phase fluctuations or a parametric decay of amplitudons into phase modes, as previously proposed \cite{liu_possible_2013}.

\section{Conclusions}

The impact of fluctuations on symmetry breaking transitions has long been considered, for example in the Peierls instability \cite{lee_fluctuation_1973, mckenzie_effect_1992, degiorgi_optical_1994, monien_exact_2001}. Providing a time-domain view of the structural relaxation pathways, the present measurements highlight the impact of long-lived structural non-equilibrium to the order parameter.

The general mechanism of amplitude suppression by CDW-coupled modes should apply also to other phases and systems. Indeed, measurements in the NC phase feature a similar behavior as the IC phase (Appendix \ref{sec:Data for nearly-commensurate (NC) phase}). Despite differences in symmetry, CDW wavevectors and electronic gaps, both phases exhibit closely related amplitude and phase excitations, as pointed out by Nakanishi and Shiba \cite{nakanishi_domain-like_1978}.

Relevant further questions pertain to the possible mechanisms of generating phasons and dislocation-type topological defects, as well as their coupling to regular lattice modes. Also the link between fluctuation modes and the creation and relaxation of metastable states\cite{shi_ultrafast_2019, stojchevska_ultrafast_2014} the influence of partial and full commensurability in the different CDW states call for further investigation. Additional insights may be gained by investigating the ultrafast phase transitions between different CDW states\cite{vogelgesang_phase_2017, lantz_domain-size_2017, laulhe_ultrafast_2017} and the populations of amplitude and phase modes in the nascent state after transition. 

Considering methodical aspects, this work represents the first comprehensive study employing ultrafast low-energy electron diffraction with a temporal resolution of 1 ps. Future investigations using ULEED will enable a quantitative analysis of the three-dimensional structural evolution based on time- and energy-dependent diffraction. Moreover, the method is applicable to a wide variety of other surface systems and low-dimensional structures, harnessing its strengths of high momentum resolution, efficient scattering and enhanced sensitivity to lattice fluctuations.

\begin{acknowledgments}
This work was funded by the European Research Council (ERC Starting Grant `ULEED',
ID: 639119) and the Deutsche Forschungsgemeinschaft (SFB-1073, project A05). We
gratefully acknowledge insightful discussions with H. Schwoerer, B. Siwick, J. D. Axe and T. Aslanyan. We thank L. Hammer for introducing us to the dynamical LEED computations. Furthermore, we thank K. Hanff for help with sample preparation.
\end{acknowledgments}

\section*{Data availability}
The data that support the findings of this study are available from the corresponding author upon reasonable request.

\appendix

\section{Methods}
\label{sec:methods}

\subsection{Experimental details}

\begin{figure*}
    \centering
    \includegraphics{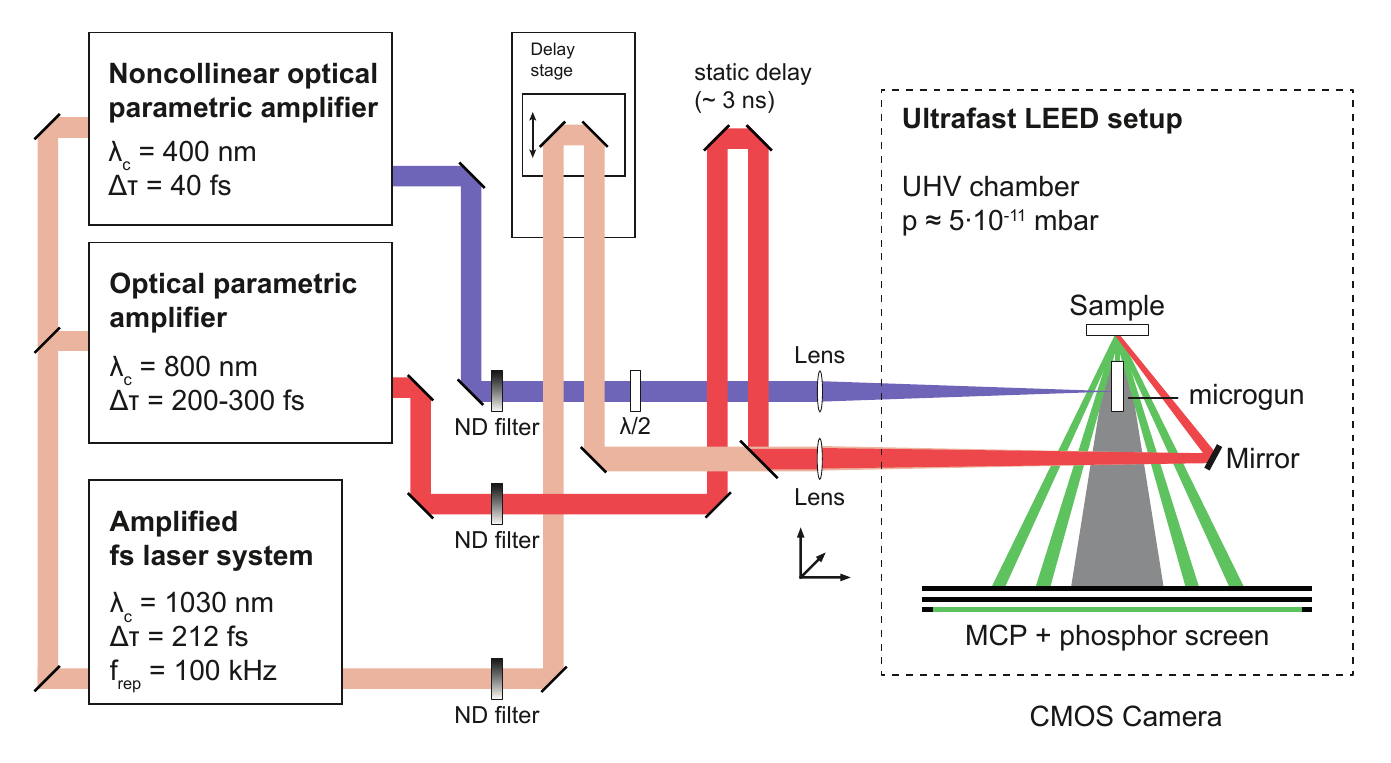}
    \caption{Schematic of the ultrafast LEED setup.}
    \label{fig:Figure07_SDY}
\end{figure*}

Here, we briefly describe our experimental ULEED apparatus (Fig.~\ref{fig:Figure07_SDY}). The time-resolved measurements are conducted in an ultra-high vacuum chamber (base pressure $p=5\cdot 10^{-11}$ mbar) into which samples are transferred via a load-lock system and cleaved in-situ. Inside the chamber, the electron source (microgun\cite{storeck_nanotip-based_2017}) and a microchannel plate detector are mounted. A cooled CMOS camera records the detected electron diffraction patterns from outside the UHV chamber.

A femtosecond laser system (Amplifier, NOPA and OPA) generates three femtosecond laser beams of different center wavelength. An ultraviolet beam (center wavelength of 400 nm) is focused on a nanometric tungsten needle that is embedded inside the microgun\cite{storeck_nanotip-based_2017} generating ultrashort electron pulses via two-photon photoemission. An electrostatic lens assembly controls the collimation of the electron beam having an energy range of 40-100~eV. With a gun front diameter of 80 $\mu$m and a working distance of around 150 $\mu$m, we achieve a temporal resolution of 1 ps and an electron beam diameter of approximately 10 $\mu$m at the sample. An upper limit of the technique’s temporal resolution is obtained by the derivative of the fastest intensity change in our data set (suppression of satellite peak at 100~eV) shown in Fig.~\ref{fig:Figure01_SDY}g. An infrared beam (1030 nm center wavelength) optically excites the sample at specific times controlled by an automated linear delay stage. In order to heat the 1\textit{T}-TaS\textsubscript{2} sample and stabilize it in the IC phase slightly above $T\approx\,353$~K, a third beam (800 nm) is aligned collinearly with the infrared beam. This pulse is delayed by about 3 ns with respect to the electron and 1030 nm pulses (i.e., it arrives 10 $\mu$s before the next pulses) and thus leads to an average increase in sample temperature.

\subsection{Data analysis}
The recorded LEED images are preprocessed to correct for minor drifts in-between measurement runs, and for distortions caused by local electromagnetic fields as well as the projection to a flat MCP detector.

In order to obtain time curves (Fig.~\ref{fig:Figure02_SDY}d) from the stacks of diffraction patterns, we process the data in a sequence of operations.  First, a binary circular mask is laid on top of each individual reflex (Fig.~\ref{fig:Figure02_SDY}c, blue and red circle, diameter $\Delta k_{\text{main}}=0.6\ \angstrom^{-1}$ and $\Delta k_{\text{sat}}=0.36 \ \angstrom^{-1}$ respectively) for each time delay. Second, we fit 2D Cauchy distributions (background: slope and constant offset) to the satellite reflexes and 2D pseudo Voigt profiles (background: slope and constant offset) to the main lattice reflexes, to determine a background profile and subtract it from each spot:

\begin{widetext}
\begin{align}
    &C(x,y)= \frac{A}{2\pi\sigma_1 \sigma_2}\cdot \left(1+\left(\frac{x}{\sigma_1}\right)^2+\left(\frac{y}{\sigma_2}\right)^2\ \right)^{-3/2}+(a\cdot x+b\cdot y)\ +C,\\
    &PV(x,y)=A\cdot\left(1+\left(\frac{x}{\sigma_1}\right)^2+\left(\frac{y}{\sigma_2}\right)^2\ \right)^{-3/2}+B\cdot e^{-\left(x/\sigma_3\right)^2-\left(x/\sigma_4\right)^2}+(a\cdot x+b\cdot y)+C.
\end{align}
\end{widetext}

Here, the x and y axes correspond to the azimuthal and radial directions (with respect to the main peak) for a given spot. Third, from the background-corrected segment, the average and the maximum intensity (average over brightest 4\% within a mask) are determined for each reflex within the mask. The remaining intensity outside the circular masks forms the integrated background. For improved signal-to-noise ratio, several spot curves are averaged, i.e the satellite curves represent the mean of the 11 brightest  reflexes. Furthermore, from the 2D fit functions, we obtain the azimuthal ($\sigma_1$) and radial ($\sigma_2$) widths for each reflex (Fig.~\ref{fig:Figure04_SDY}b).

\subsubsection{Debye-Waller-corrected amplitude signal}

Next, we describe the separation of the amplitude-quench-related intensity changes from Debye-Waller-type peak suppression for main lattice (Fig.~\ref{fig:Figure03_SDY}d) and satellite reflexes (Fig.~\ref{fig:Figure04_SDY}b). 

The dynamical LEED simulations indicate that there can be considerable differences in the coefficients $c_\textbf{s}$ (Eq. \ref{appb}). Empirically, we find that for the IC phase, the \{(0 1), (-1 0), (1 -1)\} peaks show a negligible influence of the initial quench ($c_\textbf{s}\approx 0$). In Fig.~\ref{fig:Figure03_SDY}a, for each fluence, light blue ($c_\textbf{s}\approx 0$) and dark blue curves ($c_\textbf{s}>0$) are averaged ($I_{\text{non-amp}}$ and $I_{\text{amp}}$). We now use the two peak groups to extract the temporal evolution of $A$ by removing the time-dependent Debye-Waller suppression $e^{-2W_\textbf{s}}$ from the intensity of the peaks sensitive to the PLD, with a constant factor $C_1=0.81$ that accounts for the slightly different $W_\textbf{s}$ of these peaks:

\begin{equation}
    I_{\text{ratio, main},\, F}=\left(\frac{ I_{\text{amp},F} }{I_{\text{non-amp},F_1}}\right)^{\frac{F}{F_1}\cdot C_1} = 1-c_\textbf{s} A^2
\end{equation}

The value of $C_1$ was determined by the main peak suppression at long delays (beyond 1 ns) and the lowest fluence ($F_1=1.3$\ mJ/cm$^2$), for which a negligible amplitude change is expected. The value of $c_\textbf{s}$ for the amplitude-sensitive peaks is determined from the $A^2$-intensity dependence of the phonon-corrected satellite peaks (Eq. \ref{appb}), evaluated at maximum suppression of the lowest fluence.

From the satellite reflexes (11 brightest spots), the corrected amplitude is obtained similarly using the Debye-Waller-dominated main lattice curve $I_{\text{non-amp},F_1}$ for the lowest fluence with the factor $C_2=1.2$

\begin{equation}
    I_{\text{ratio, sat},\, F}=\left(\frac{ I_{\text{sat},F} }{I_{\text{non-amp},F_1}}\right)^{\frac{F}{F_1}\cdot C_2} = A^2.
\end{equation}
\\
\subsubsection{Fitting of time constants}
The fit function in Fig.~\ref{fig:Figure03_SDY}d is based on a step-like decrease followed by two exponential relaxations 

\begin{widetext}
\begin{equation}
   S(t)=1-\theta(t-t_0) \cdot (-A_1 + A_2\cdot (1-e^{-(t-t_0)/\tau}))+A_3\cdot(1-e^{-(t-t_0)/\tau_2})
\end{equation}
\end{widetext}

where $\theta$ is the Heaviside function, $t_0$ time zero, $A_1$, $A_2$ and $A_3$ are the amplitudes and $\tau_1$ and $\tau_2$ time constants. The complete fit function is the convolution of $S(t)$ with a Gaussian (FWHM of 1 ps) corresponding to the temporal resolution in our experiment.

\section{Data for nearly-commensurate (NC) phase}
\label{sec:Data for nearly-commensurate (NC) phase}

\begin{figure*}[ht!]
    \centering
    \includegraphics{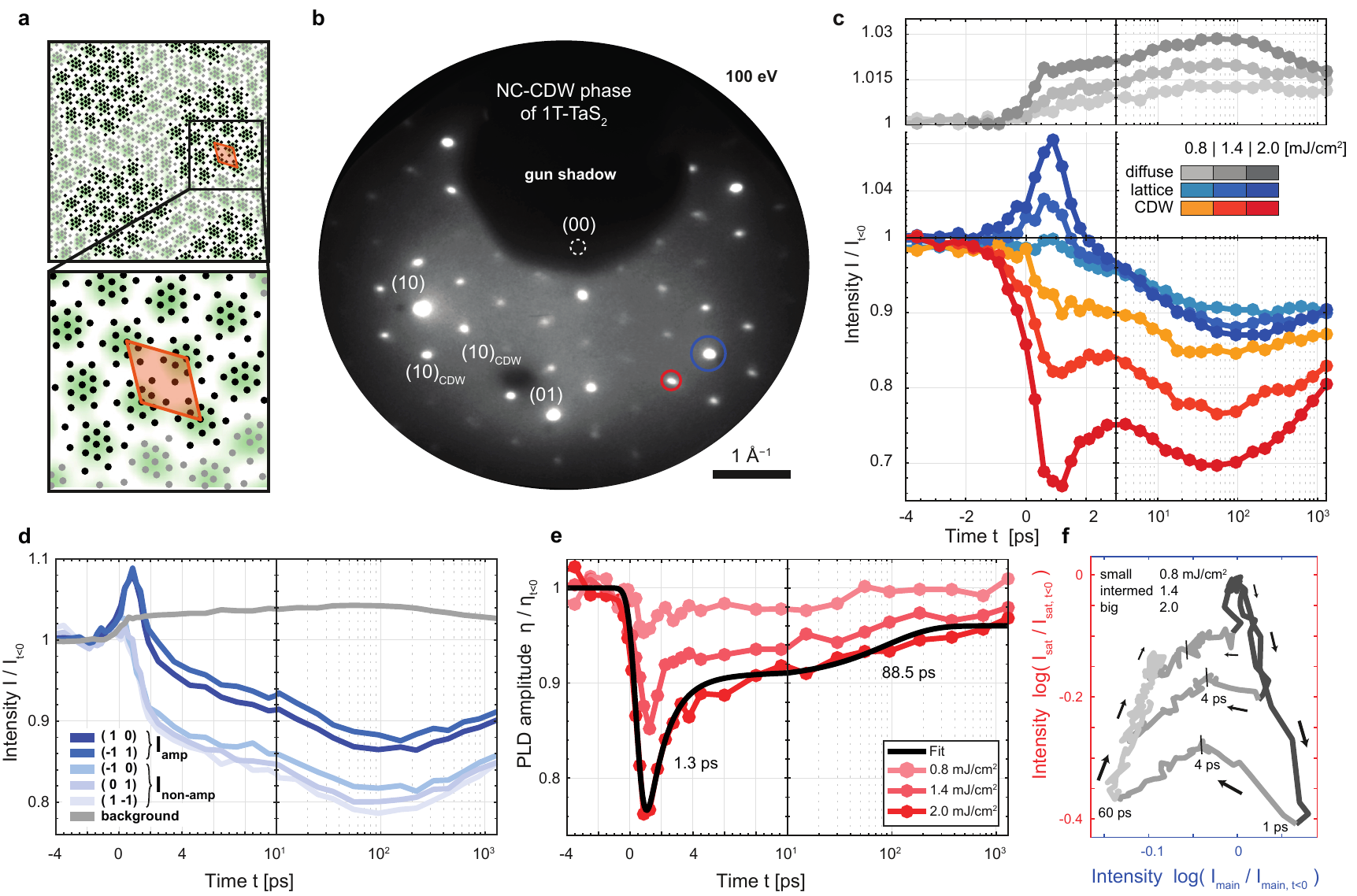}
    \caption{Measurements in the NC phase for an electron energy of 100~eV. (a) Top view of nearly-commensurate (NC) CDW phase illustrating charge density (green), distorted lattice (black dots, displacements exaggerated) and superstructure unit cell (orange). (b) Diffraction pattern of the NC phase of 1\textit{T}-TaS\textsubscript{2} showing main lattice reflexes and several orders of PLD induced satellites (integration time: 90 s). (c) Time-dependent measurement of reflexes (blue and red circles in (b)) and diffuse background (for three fluences). The main lattice signal is averaged over the (10) and (-1 1) spots (blue), the satellite signal over several reflexes. Curves are normalized to signal at negative times. (d) Time-dependent intensity of visible main lattice reflexes and integrated background intensity, for a fluence of $F=2.0\,$mJ/cm$^2$. Two inequivalent classes of spot groups are found, featuring a strong (dark blue) and weak (light blue) sensitivity to the amplitude quench. (e) Extracted PLD amplitude quench and relaxation for three fluences, showing a rapid and a slower relaxation component. (Time constants from a biexponential fit (black line) to the highest fluence data: 1.3 ps and 88.5 ps). (f) Main lattice peak intensities vs. satellite peak intensities, leading to cyclic trajectories in a 2D plane with varying size. Following a two-stage relaxation, all curves reach a common equilibrium line after approximately 60 ps. The gray color scale highlights certain time intervals (dark gray: 0-1 ps, intermed. gray: 1-60 ps, light gray: 60-1500 ps).}
    \label{fig:Figure08_SDY}
\end{figure*}

Figure \ref{fig:Figure08_SDY} displays the analysis discussed above applied to the nearly-commensurate phase. Similar features are found in the pump-probe curves for the main and satellite diffraction peaks as well as the background (Fig.~\ref{fig:Figure08_SDY}c), the long-lived amplitude suppression (Fig.~\ref{fig:Figure08_SDY}d-e) and the relaxation cycles (Fig.~\ref{fig:Figure08_SDY}f). In particular, the two-stage amplitude relaxation process (first stage up to ~4 ps, second stage up to ~60 ps) is very pronounced at all fluences.

\section{Data at 80~eV electron energy}
\label{sec:Data at 80eV electron energy}

\begin{figure*}[ht!]
    \centering
    \includegraphics{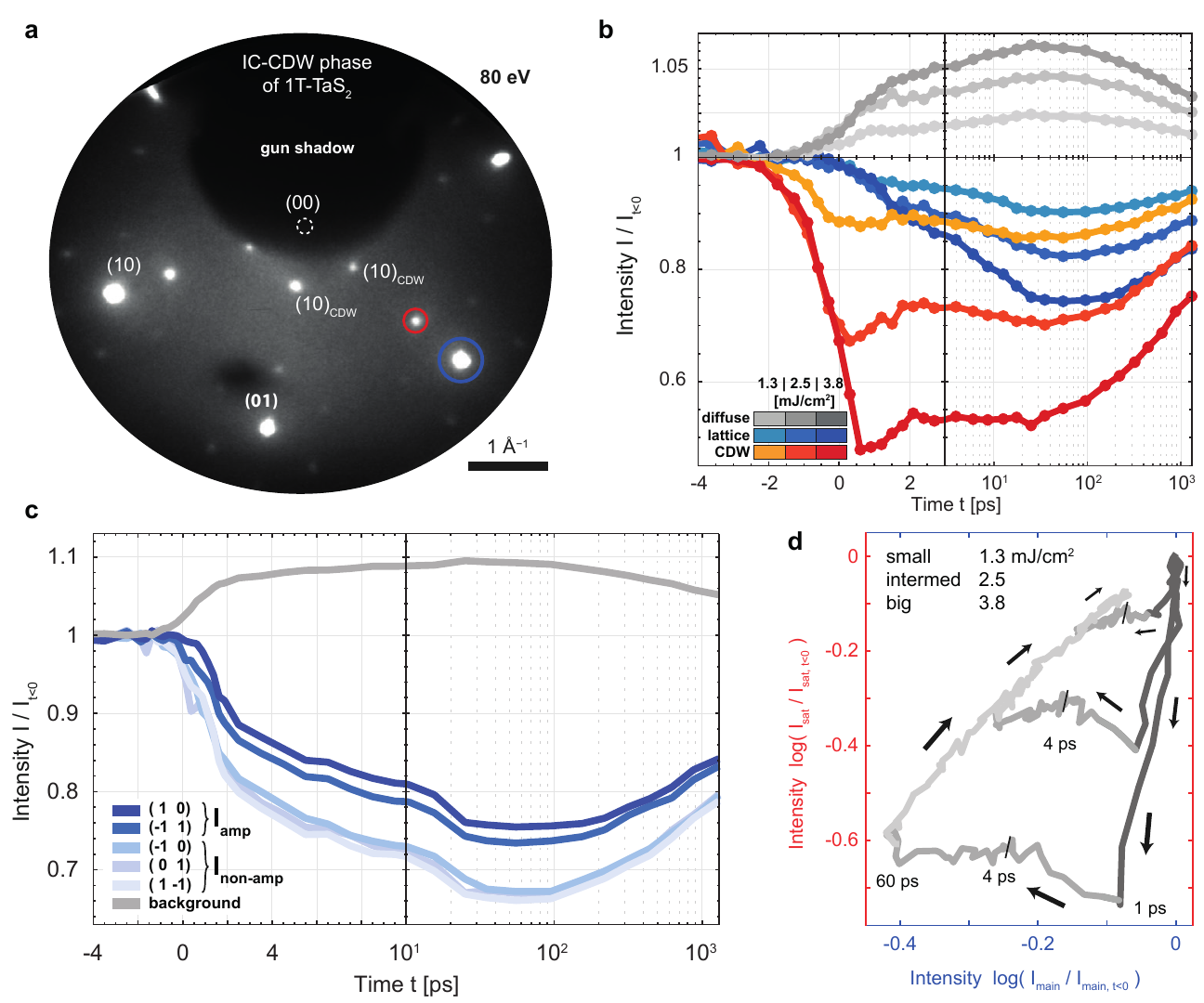}
    \caption{Measurements in the IC phase for an electron energy of 80~eV. (a) Diffraction pattern of the IC phase of 1\textit{T}-TaS\textsubscript{2} showing main lattice reflexes and first-order PLD-induced satellites (integration time: 90 s). (b) Time-dependent measurement of reflexes (blue and red circles in (a)) and diffuse background (for three fluences). The main lattice signal is averaged over the (10) and (-1 1) spots (blue), the satellite signal over several reflexes. Curves are normalized to signal at negative times. (c) Time-dependent intensity of visible main lattice reflexes and integrated background intensity, for a fluence of $F=3.8$ mJ/cm$^2$. Two inequivalent classes of spot groups are found, but none shows a strong dependence on the amplitude quench. (d) Main lattice intensity vs. satellite peak intensity, leading to cyclic trajectories in a 2D plane with varying size. Note that all curves reach a common equilibrium line after approximately 60 ps. The gray color scale highlights certain time intervals (dark gray: 0-1 ps, intermed. gray: 1-60 ps, light gray: 60-1500 ps).}
    \label{fig:Figure09_SDY}
\end{figure*}

Figure \ref{fig:Figure09_SDY} shows additional data recorded in the IC phase at 80~eV energy. The main lattice peaks show a much weaker dependency on the PLD amplitude (Fig.~\ref{fig:Figure09_SDY}b-d).

\section{Relaxation cycles for main peaks (-1 1), (0 1) and (1 -1)}
\label{sec:Relaxation cycles for main peaks}

\begin{figure*}[ht!]
    \centering
    \includegraphics{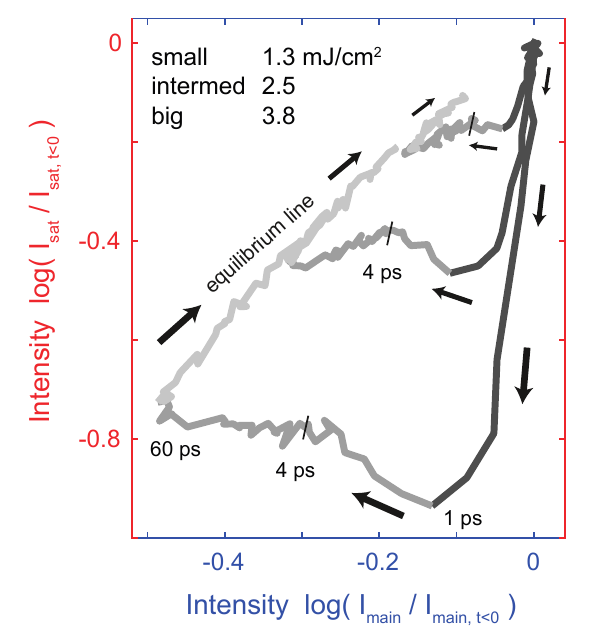}
    \caption{Main lattice peaks without amplitude feature vs. satellite peak intensities, leading to cyclic trajectories in a 2D plane with varying size. All curves reach a common equilibrium line after approximately 60 ps. The gray color scale highlights certain time intervals (dark gray: 0-1 ps, intermed. gray: 1-60 ps, light gray: 60-1500 ps).}
    \label{fig:Figure10_SDY}
\end{figure*}

Figure \ref{fig:Figure10_SDY} shows the relaxation cycles in the IC phase as in Fig.~\ref{fig:Figure05_SDY}a, using the intensities of the main lattice peaks (-1 1), (0 1) and (1 -1) without sensitivity to the PLD amplitude.

\section{Impact of CDW defects on peak width}
\label{sec:Impact of topological defects on peak width}

Here, we argue that our data rules out a linear scaling of CDW defect density with fluence and is only consistent with a nonlinear or threshold behavior. Assuming a linear relation of the defect density with the fluence $n\sim F$ and a correlation length $L\sim 1/\sqrt{n}$ (see Ref. \cite{vogelgesang_phase_2017}), the defect-induced broadening should scale as $\sigma_{top}\sim 1/L \sim \sqrt{n} \sim \sqrt{F}$.  A doubling of the normalized peak widths $\sigma_{tot}=\sqrt{{\sigma_0}^2+{\sigma_{top}}^2}$ with respect to the instrument resolution $\sigma_0$ for the highest fluence would then imply considerably higher broadening values for lower fluences ($\sigma_{tot,\ 2.5}=1.7$ and $\sigma_{tot,\ 1.3}=1.4$) than observed in our measurement (Fig.~\ref{fig:Figure04_SDY}c). Experimentally, we find maximum broadening values at $t\approx 1$\ ps of $\sigma_{tot,\ 3.8}\approx 2$\ , $\sigma_{tot,\ 2.5} \approx 1.3$\ and $\sigma_{tot,\ 1.3}\approx 1$\ for the highest, intermediate and lowest fluence, respectively (see Fig.~\ref{fig:Figure04_SDY}c). Thus, we infer that the density of topological defects does not scale linearly with fluence.

\section{Dynamical LEED computation}
\label{sec:Dynamical LEED computation}

We performed dynamical LEED simulations on the commensurate CDW phase of 1\textit{T}-TaS\textsubscript{2}, varying the atomic displacements of sulfur and tantalum continuously from the undistorted structure towards the C-phase structure recently reconstructed \cite{von_witte_surface_2019}. We are aware that the C phase is a simple approximation for the description of the high-temperature incommensurate CDW phase. However, it exhibits the same crucial feature of opposing sulfur displacements that we believe is responsible for the different sensitivities of the main lattice peaks. Also, dynamic LEED calculations involve high computational effort, in particular for large unit cells necessary for incommensurate structures. The obtained data contains PLD-amplitude- and energy-dependent scattering intensities for main lattice and CDW satellites spots. In the following, we focus on main lattice diffraction intensities.

In the electron energy range of 70-140~eV, the diffraction intensity is mainly determined by scattering from sulfur atoms, explaining the strong dependence from the PLD amplitude of sulfur atoms (Fig.~\ref{fig:Figure01_SDY}1a). 

\begin{figure*}[ht!]
    \centering
    \includegraphics{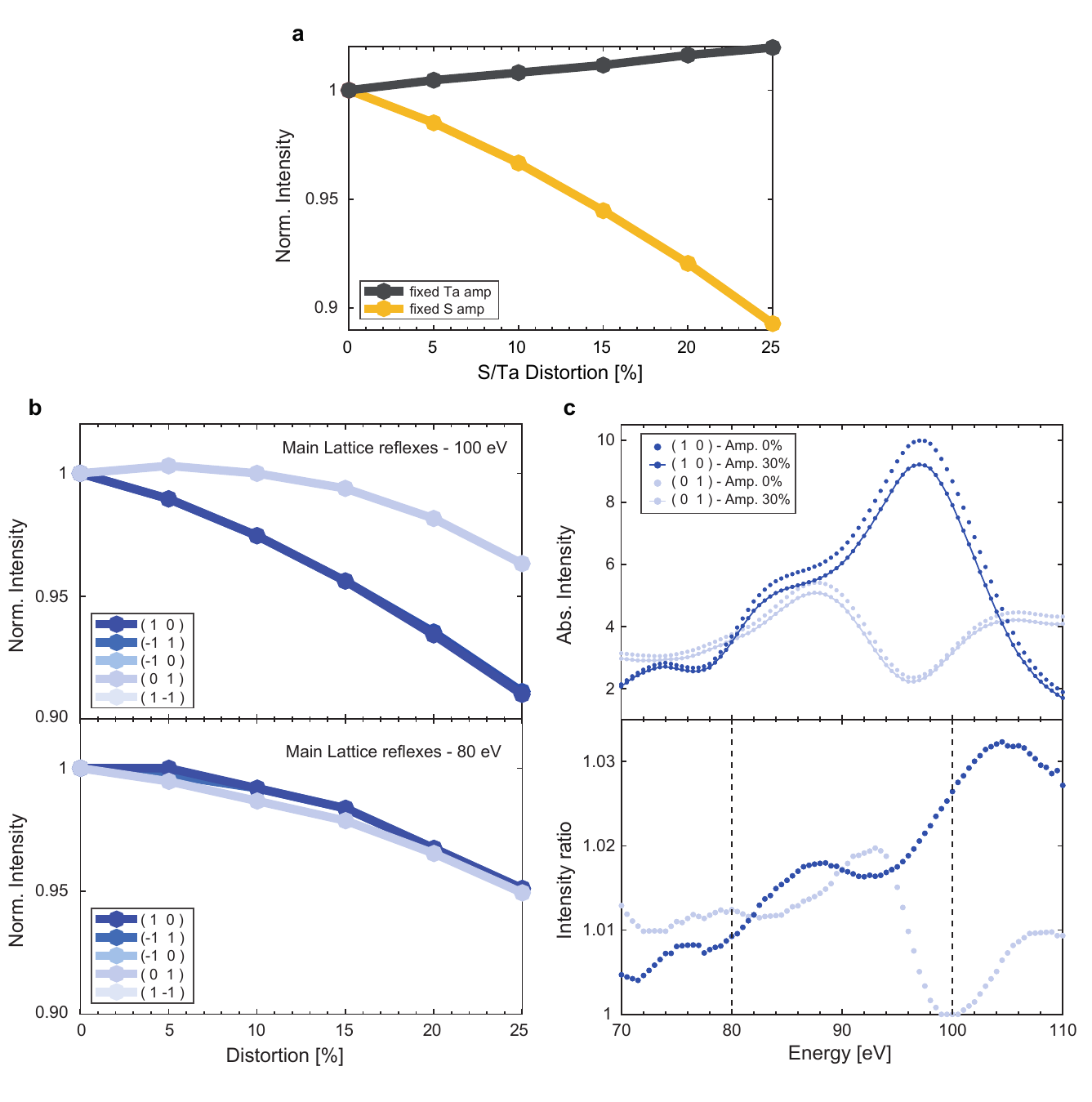}
    \caption{Dynamical LEED simulations. (a) Normalized intensity of main lattice reflex (1 0) as a function of sulfur and tantalum displacement for an electron energy of 100~eV. Enhanced scattering off sulfur atoms results in a much stronger dependence on the sulfur atom displacements. (b) Normalized intensities of main lattice spots for an electron energy of 80 and 100~eV as a function of the fraction of the maximum commensurate PLD amplitude. The diffraction reflexes split up into two spot groups. Light and dark blue curves coincide, respectively, due to normal incidence of the electron beam. (c) LEED spectra (top) for both groups (light and dark blue) for vanishing (points) and finite (dash points) distortion. The percentage refers to the amplitude of the commensurate PLD in low-temperature phase. The intensity ratio (bottom) illustrates the energy-dependent sensitivity between reflex groups.}
    \label{fig:Figure11_SDY}
\end{figure*}

Figure \ref{fig:Figure11_SDY}b shows PLD dependent intensities for an electron energy of 100 and 80~eV, each normalized to the intensity value for zero distortion (metal structure). The PLD amplitude range is adapted to the expected values realized in the incommensurate phase \cite{scruby_role_1975} which are assumed to be considerably smaller ($\sim 30\%$ of PLD amplitude of the commensurate low-temperature phase). In this range for 100~eV, we can show that there are two groups of main lattice spots that respond differently upon PLD change, whereas for 80~eV, all intensities follow a common curve. Moreover, the magnitude of the relative intensity changes approximately matches the observed ones in the experiment. The curves within a group of main lattice peaks \{(1 0), (-1 1)\} and \{(-1 0), (0 1), (1 -1)\} conincide since the simulation is performed at normal incidence.

Figure \ref{fig:Figure11_SDY}c shows energy-dependent intensity curves for two main lattice peaks contained in one of the two groups (light and dark blue), each for zero PLD and 30\% PLD amplitude of the commensurate low-temperature phase. The ratio of spectra for each spot with minimal and maximal amplitude (Fig.~\ref{fig:Figure11_SDY}c, bottom) displays a rich oscillatory behavior. Importantly, however, for energies of 80~eV and 100~eV, the spots exhibit a drastically different sensitivity to PLD changes, with a small and large difference for the separate spot groups, respectively. These predictions directly corroborate our experimental findings at the different electron beam energies.

\clearpage
\input{main.bbl}

\clearpage

\end{document}

%% file: main.bbl
%